\documentclass[useAMS,usenatbib]{mn2e}
\usepackage{graphicx}
\usepackage{longtable}
\newcommand{\msol}{$M_{\odot}$}

\title
[The Population of Compact Star Clusters in M81]
{Wide-field HST/ACS images of M81: The Population of Compact Star Clusters}
\author[M. Santiago-Cort\'es, Y. D. Mayya and
  D. Rosa-Gonz\'alez]{M. Santiago-Cort\'es$^{1}$\thanks{E-mail: scortes@inaoep.mx; ydm@inaoep.mx; danrosa@inaoep.mx},
Y. D. Mayya$^{1}$\footnotemark[1] 
and D. Rosa-Gonz\'alez$^{1}$\footnotemark[1]\\
$^{1}$Instituto Nacional de Astrof\'isica \'Optica y Electr\'onica, Luis Enrique Erro 1, Tonantzintla 72840, Puebla, Mexico}

\begin{document}
\date{Accepted 2010 February 15.  Received 2010 February 9; in original form 2009 September 11
}
\pagerange{\pageref{firstpage}--\pageref{lastpage}} \pubyear{2009}

\maketitle

\label{firstpage}

\begin{abstract}
We study the population of compact stellar clusters (CSCs) in M81, using the 
\textit{HST/ACS} images in the filters F435W, F606W and F814W covering, 
for the first time, the entire optical extent of the galaxy. Our sample
contains 435 clusters of FWHM less than 10~ACS pixels (9~pc). 
The sample shows the presence of two cluster populations, a blue 
group of 263 objects brighter than $B=22$~mag, and a red group 
of 172 objects, brighter than $B=24$~mag. 
Based on the analysis of colour magnitude diagrams and making use of simple
stellar population models, we find the blue clusters are younger than
300~Myr with some clusters as young as few Myr, and the red clusters are as old as
globular clusters.
The luminosity function of the blue group follows a power-law distribution 
with an index of $2.0$, typical value for young CSCs in other galaxies. 
The power-law shows unmistakable signs of truncation at $I=18.0$~mag 
($M_I=-9.8$~mag), which would correspond to a mass-limit of 
$4\times 10^{4}\,M_{\odot}$ if the brightest clusters are younger than 10~Myr.
The red clusters have photometric masses between $10^{5}$ to $2\times 10^{7}\,M_{\odot}$ for 
the adopted age of 5~Gyr and their luminosity function resembles very much the 
globular cluster luminosity function in the Milky Way.
The brightest GC in M81 has $M_B^0=-10.3$~mag, which is $\sim0.9$~mag 
brighter than $\omega$~Cen, the most massive GC in the Milky Way.
\end{abstract}

\begin{keywords}
catalogs -- galaxies: individual (M81) -- galaxies: spiral -- galaxies: star clusters
\end{keywords}

\section{INTRODUCTION}
With the advent of the \textit{HST (Hubble Space Telescope)} a new class of
stellar clusters have been identified: the Compact Star Clusters (CSCs) with typical
masses of $\sim10^{4}$ to $10^{6}\,M_{\odot}$ and sizes between 1 and
6 pc \citep{Meurer1995}. CSCs have been found in several environments,
including violent star forming regions within interacting galaxies
\citep{Whitmore1999}. 
The similarity between the compactness and mass of the CSCs and that of the
globular clusters (GC) is a reason to think of an evolutionary connection
between them.
Moreover, the compact stellar clusters are unique laboratories for studying
diverse star formation processes related to the star formation history of 
the host galaxy. 
The detailed studies of globular clusters --- with ages comparable to the age
of the Universe --- have revealed the early formation history of nearby
galaxies and the Milky Way \citep{Harris1996, Barmby2003}, whereas the
studies of younger CSCs --- ages $<1$~Gyr --- delineate the recent star
formation history, that in some cases are related to interactions 
with neighbouring galaxies \citep{Holtzman1992, Whitmore1999, Mayya2008}.

M81 (NGC 3031) is a large Sab spiral galaxy, very similar to M31 in appearance
and roughly as massive as the Milky Way (MW). 
M81 at a distance of 3.63 Mpc [$m-M=27.8\pm0.2$; \cite{Freedman1994}] is the
biggest member of the M81 Group, which includes the prototype starburst
galaxy M82. An interaction $\sim$100--500 Myr ago between different members 
of this group has been discussed by several authors 
\citep{Brouillet1991, Yun1999}. 
Recent observations of M82 show that it has a large population of CSCs, with
young clusters (age $<10$~Myr) concentrated towards the center, and 
relatively older clusters ($\sim100$~Myr) homogeneously distributed across the
disk, the latter population having formed as a part of the disk-wide burst
following the interaction \citep{Mayya2006,Mayya2008,2009ApJ...701.1015K}. It is of interest to 
investigate whether the interaction also triggered CSC formation in M81.

The population of GCs in M81 has been studied in the past by several groups.
\citet{PerelmuterR1995} used an extensive database to find $\sim70$ objects
classified as cluster candidates in the inner 11~kpc radius of M81.
After completeness corrections for the  unobserved area, they estimated the total 
GC population would be $210\pm30$. \citet{PerelmuterBH1995} obtained spectra
of 82 cluster candidates and confirmed 25 as bona fide globular clusters. 
The derived mean metallicity of the globular clusters was  
$[Fe/H]=-1.48\pm0.19$ confirming previous results from~\citet{Brodie1991}. 
\citet{Schroder2002} obtained spectra of 16 additional globular cluster
candidates selected from an extended list of \citet{PerelmuterR1995} catalogue 
and confirmed all of them to be GCs. Hence, in total there are 41 objects
that are confirmed as globular clusters using spectroscopic data. 
The metallicity distribution of these GCs 
is similar to that in M31 and the Milky Way, two galaxies that are 
morphologically very similar to M81. Hence, it of interest to determine 
whether M81 also contains a similar number of total GCs as in the Milky Way.

\citet{Chandar2001} and \citet{Chandar2001b} carried out a search for compact objects in M81 based on
observations with the \textit{HST/WFPC2} camera. They discovered 114 CSCs
in an area of $40\,arcmin^{2}$. The analysis found, for the first time,  
two different cluster populations, 59 red clusters [$(B-I)_{0}\geq0.85$ mag] 
which are candidate for globular clusters and 55  young clusters with 
photometric ages  $<600$~Myr. The authors related the 
latter population with the interaction between M81 and M82.

In the present work, we carried out a search for CSCs in 29 adjacent HST/ACS 
fields centered on the nucleus of M81. The present dataset offers not only an 
improved spatial resolution, but also covers a field of view that is 8.5 
times larger than that covered by \citet{Chandar2001b}. 
Results obtained from a subset of 12 adjacent central 
fields were presented in \citet{2009arXiv0903.1619S}.

This paper is organized as follows: \textsection 2 presents the observational 
material used in this work; \textsection 3 gives a summary of the cluster detection
and selection method; \textsection 4 describes the analysis of colour magnitude 
diagrams (CMD) and luminosity functions (LF) for the selected clusters; 
the discussion and conclusions of these studies are presented in \textsection 5.

\section{OBSERVATIONS}
\begin{figure}
 \centering
  \includegraphics[width=\columnwidth]{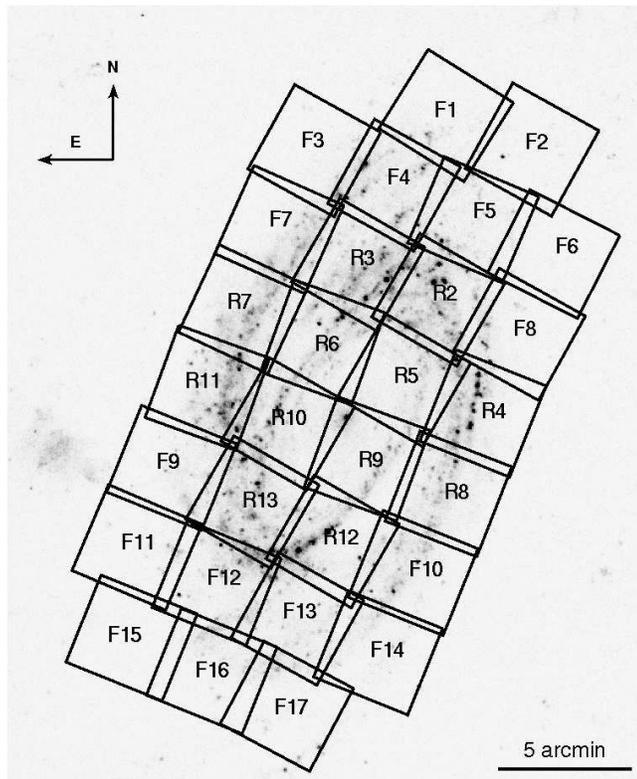}%{Fig1}
  %\vskip -3mm
\caption{The footprints of 29 \textit{HST/ACS} pointings superposed on
a $23'\times28'$ \textit{GALEX} image 
of M81. Identification number of each field is indicated. These 29 pointings 
cover the entire optical/UV extent of the galaxy.} 
  \label{fig:simple}
\end{figure}

The observations used in this work were carried out with the ACS Wide Field 
Channel on board the \textit{HST}. They were part of the projects with proposal
IDs 10250 (PI: John Huchra) and 10584 (PI: Andreas Zezas). 
Table~\ref{tab:filters} lists the details of the observations. The  \textit{HST} 
database contains 29 adjacent fields covering a field of view 
of $\sim340\,arcmin^{2}$ (See Figure~\ref{fig:simple}), with a sampling
of $0.05^{\prime\prime}{\rm pix}^{-1}$ (0.88~pc~${\rm pix}^{-1}$).
For each field, observations were carried out in the F435W, F606W 
and F814W filters, which for the sake of brevity we will refer to as
$B$, $V$ and $I$ filters, respectively, throughout this paper. 
The standard pipeline process (CALACS) provided by the 
Hubble Heritage Team were used for bias, dark and
flat-field corrections.  The pipeline uses
the IRAF/STSDAS Multidrizzle task to combine the images of a single field and produces
weight maps related with the background and instrumental noise. Also this task 
corrects bad pixels, rejects cosmic rays, and eliminates artifacts \citep{Mutchler2007}.
However, images taken with different programs have slightly different astrometric
coordinates. We used common stars in the adjacent images to tie all the images to a single
coordinate system.

\begin{table}
\caption{Filters and exposure times.}
\begin{center}
{\small
\begin{tabular}{cccc}
\hline
Field ID & Filter& Proposal ID   &  Exp. Time (s)\\

\hline
\hline
F1         &  F435W   &    10584&   $1 \times  900$   \\
F1         &  F606W   &    10584&   $1 \times  880$   \\
F1         &  F814W   &    10584&   $1 \times  895$   \\
F2         &  F435W   &    10584&   $3 \times 1565$   \\
F2         &  F606W   &    10584&   $3 \times 1580$   \\
F2         &  F814W   &    10584&   $3 \times 1595$   \\
F3--F14    &  F435W   &    10584&   $2 \times 1200$   \\
F3--F14    &  F606W   &    10584&   $2 \times 1200$   \\
F3--F10    &  F814W   &    10250&   $3 \times 1650$   \\
F11        &  F814W   &    10250&   $2 \times 1100$   \\
F12--F14   &  F814W   &    10250&   $3 \times 1650$   \\
F15--F16   &  F435W   &    10584&   $3 \times 1565$   \\
F15--F16   &  F606W   &    10584&   $3 \times 1580$   \\
F15--F16   &  F814W   &    10584&   $3 \times 1595$   \\
F17        &  F435W   &    10584&   $2 \times  665$   \\
F17        &  F606W   &    10584&   $1 \times  350$   \\
F17        &  F814W   &    10584&   $1 \times  350$   \\
R2-R13     &  F435W   &    10584&   $2 \times 1200$   \\
R2-R13     &  F606W   &    10584&   $2 \times 1200$   \\
R2-R13     &  F814W   &    10250&   $3 \times 1650$   \\

\hline
\end{tabular}\newline\newline
}
\end{center}\label{tab:filters}
\end{table}

\begin{table*}
\caption{HST/ACS point sources erroneously classified as globular clusters in previous studies\label{tab:noGCs}}
\begin{minipage}{\textwidth}
\renewcommand{\thefootnote}{\thempfootnote}
{\small
\begin{tabular*}{\textwidth}{cccccccrr}
\hline
ID  &    RA(2000) &  Dec(2000) & $B$ (mag) & $B-I$& FWHM(pix) & ID$^{b}$ & $V_{\rm rad}^{b}$ (km/s) & [Fe/H]$^{b}$ \\
\hline
\hline
%   ID   AR[hh:mm:ss]  Dec[gg:mm:ss]  B [mag]   B-I      FWHM   Other_ID_PBH1995(a)    v[Km/s][Fe/H]   
 45861F1  & 09:55:03.823 & 69:15:38.10  &  19.09  &   1.33 &   2.27$^a$   &  Is40165  & $   6$   & $-1.57$    \\
  9514F3  & 09:55:44.079 & 69:14:12.00  &  20.11  &   2.14 &   1.74$^a$   &  Is40181  & $  46$   & $ 0.64$    \\
   113F10 & 09:55:06.265 & 68:56:24.78  &  18.60  &   1.00 &   2.47$^a$   &  Is50037  & $ -18$   & $-2.34$     \\
  8041F11 & 09:56:40.582 & 68:59:52.44  &  19.68  &   1.95 &   2.61$^a$   &  Is50225  & $  -7$   & $-0.04$     \\
  3740F6  & 09:54:19.980 & 69:09:11.57  &  19.97  &   1.45 &   2.02$^a$   &  Is51027  & $ 300$   & $-2.47$      \\
   901F17 & 09:55:56.866 & 68:52:13.42  &  19.63  &   1.60 &   1.85$^a$   &  Is60045  & $ -28$   & $-1.03$       \\
 10870R8  & 09:54:58.754 & 69:00:58.21  &  20.97  &   1.72 &   2.19       &  Is50286  & $  -9$   & $-0.04$       \\
 10600F9  & 09:56:31.774 & 69:02:38.47  &  21.38  &   2.20 &   2.15       &  Id50401  & $-283$   & $-0.04$        \\
\hline 
\end{tabular*}
Note:  (a) mildly saturated stars, (b) last 3 column data were taken from \citet{PerelmuterBH1995}.
}
\end{minipage}
\end{table*}

\begin{table*}
\caption{Observational properties of compact stellar clusters of both blue and red groups$^a$\label{tab:catalog}}
\begin{minipage}{\textwidth}
\renewcommand{\thefootnote}{\thempfootnote}
{\small
\begin{tabular*}{\textwidth}{lcccrrccc}
\hline
ID$^b$  &    RA(2000) &  Dec(2000) & $B_{\rm iso}^c$ (mag) & $B-I^{d}$ & $B-V^{d}$ & $B_{\rm aper}^c$ & FWHM (pix) & $\epsilon$ \\
\hline
\hline
    R05R06584 & 148.8418408 &  69.1105121 & 17.829 &  2.058 &  1.103 & 18.185 &  3.62 &  0.03  \\
    R13R13715 & 149.1149429 &  69.0194455 & 18.437 &  1.777 &  1.000 & 18.726 &  5.82 &  0.13  \\
    R04B15666 & 148.6774430 &  69.0606742 & 18.448 &  0.317 &$-$0.067& 18.652 &  4.41 &  0.28  \\
    R10R03509 & 148.9172358 &  69.0695361 & 18.595 &  1.811 &  1.020 & 18.797 &  7.44 &  0.04  \\
    R10R10692 & 149.0356195 &  69.0642516 & 18.758 &  2.164 &  1.204 & 19.118 &  3.65 &  0.15  \\
    R03B16992 & 148.8193827 &  69.1487218 & 18.762 &  0.146 &  0.118 & 19.071 &  3.62 &  0.19  \\
    R10R09559 & 148.9417949 &  69.0501834 & 18.771 &  2.240 &  1.267 & 18.996 &  3.45 &  0.12  \\
    R06R14272 & 148.8788369 &  69.1274761 & 19.050 &  2.054 &  1.160 & 19.317 &  3.39 &  0.12  \\
    R05R06792 & 148.7596607 &  69.0938767 & 19.068 &  1.996 &  1.137 & 19.328 &  5.34 &  0.09  \\
    R05R10583 & 148.8424549 &  69.0886357 & 19.140 &  2.122 &  1.228 & 19.257 &  3.74 &  0.02  \\
    R12B14433 & 148.9722627 &  68.9844062 & 19.258 &  0.282 &  0.137 & 19.305 &  3.02 &  0.36  \\
    R02B09480 & 148.7397628 &  69.1468477 & 19.309 &  0.001 &$-$0.053& 19.404 &  9.58 &  0.35  \\
    F14B08146 & 148.8935116 &  68.9303161 & 19.326 &  1.095 &  0.465 & 19.417 &  3.51 &  0.22  \\
    R04B12769 & 148.7003380 &  69.0560965 & 19.377 &  0.619 &  0.973 & 19.986 &  5.68 &  0.31  \\
    R13R19709 & 149.0469244 &  68.9835085 & 19.381 &  2.280 &  1.361 & 19.399 &  2.72 &  0.10  \\
\hline
\end{tabular*}
Note:  (a) Only the brightest 15 (in band B) are given here. A complete list 
is available in electronic version of the Journal. 
(b) The first three digits of the ID indicate the ACS field number (Figure~\ref{fig:simple}),
the fourth letter indicates whether the cluster belongs to the blue (B) or 
red (R) group, and the remaining part is the SExtractor identification number of the source.
(c) $B_{\rm iso}$ is the ISOPHOT magnitude and $B_{\rm aper}$ is
the aperture magnitude within a radius of 10 pixel ($0.5^{\prime\prime}$) 
as defined in SExtractor. 
(d) $B-I=F435W-F814W$ and $B-V=F435W-F606W$ are the Vega system colours in the 
HST filters $F435W$, $F606W$ and $F814W$.  The colours are calculated using 
the aperture magnitudes within a radius of 10 pixel.
}
\end{minipage}
\end{table*}

\section{CLUSTER DETECTION  AND SELECTION}
We used the automatic detection code SExtractor to create an unbiased sample of cluster candidates. 
SExtractor automatically detects sources on fits images, makes photometry
and generates a data catalog~\citep{Bertin1996}. SExtractor first 
generates a background map by computing the mean and the standard deviation of
every section of the image with a user-defined grid size for which we choose $64\time64$ pixels.
The local background is clipped iteratively 
until the values in every remaining pixel is within $\pm3\sigma$ of the median
value. The mean of the clipped histogram is then 
taken as the local background. Every area of at least five adjacent pixels
that exceeded the background 
by at least $3\sigma$ was called a source candidate.

The B band was used for the detection of candidates, and we carried out 
aperture photometry of all the detected sources in each of the B, V 
and I images. The process was repeated for each of the 29 fields, resulting
in a preliminary list of 565,438 sources. 
This list contains both unresolved (stellar-like) and
resolved (extended) objects. 
The distribution of sizes peaks at FWHM=2.1~pixel, which 
corresponds to the typical Point Spread Function (PSF) of the ACS images. 
We chose $FWHM=2.4$~pixel as the dividing line to separate cluster candidates
from point sources. Our aim is to create a catalog of 
compact sized clusters and hence we restricted our catalog to sources with 
$FWHM<10$~pixel. Thus our preliminary list of CSC candidates includes all
sources with $2.4<FWHM<10$ pixels.

Among the resolved objects selected using the above mentioned criterion, we 
have two kinds of sources that contaminate the genuine CSCs. The first of
these contaminating sources is formed by the unevenness of the local
background due to the presence of dust and complex small-scale disk structures.
The second type of contaminating sources is caused due to the 
blending of several point sources due to stellar crowding. 
These contaminating sources are rejected by using the {\it AREA} parameter 
(defined as the number of contiguous pixels above the $3\sigma$ detection 
limit) of SExtractor.
By visual inspection of the images, we found that the fraction of 
contaminating sources is highest for sources having area less than 50 pixels.
Hence, we rejected all sources if they have an $AREA<$50 pixels.
By numerical calculations, we found that even the most compact objects have
$AREA>$50 pixels if they are brighter than $B=23$~mag, which effectively sets
the completeness limit of our selection process. 

While a great majority of blended stars are eliminated from the list by
the imposed AREA criteria, some of them still sneak through. In order to
eliminate such sources, we analyzed the ellipticities of the sources.
All genuine clusters are expected to be round with ellipticity $\epsilon < 0.1$. 
However, because SExtractor measures ellipticities at the limiting ($3\sigma$) 
isophote level, we found that some genuine clusters have $\epsilon >0.1$.
This happens when a cluster is surrounded by a diffuse background or is 
immersed within a stellar group. 
This kind of source is characterized by a prominent 
peak, with the aperture photometry saturating at a small radius. On the
other hand, aperture magnitude of a source formed by an elongated chain of 
stars would continue to rise with increasing radius.
This property allows us to separate clusters from the blended stars
even when the measured ellipticities are $>0.1$. 
We found that if the difference between the  aperture magnitudes of diameters
2 and 4 pixels is less than 1.5 mag, then they are true clusters. Hence, 
among the elongated sources, we retained only those sources if the difference
in magnitudes between apertures of diameters 2 and 4  pixels is less than 1.5~mag.
In summary, all sources with $2.4<FWHM<10$~pixels, $AREA>50$~pixels and
ellipticity $<0.1$ are retained, whereas among the elongated sources, only
those showing evidence of a compact core are retained.

Note that at the distance of M81, 2.4 pixel corresponds to a physical 
scale of 2.1~pc.
Given that the PSF of ACS images is 2.1 pixels, all clusters smaller than
1~pc of FWHM will have a measured FWHM of 2.4~pixels, and hence our
method cannot recover clusters more compact than this, if present.
A Gaussian FWHM of  1~pc corresponds to a 
core radius of 0.5~pc for a King profile with c={${r_t\over{r_c}}$=30}
\citep{Chandar2001}, which is almost the limiting size for the compact 
clusters, and very few such clusters are known to exist \citep{Barmby2006, Scheep2007, Ashman2001}.
Hence, we are not missing many clusters because of this criterion.
The measured FWHM$>5$~pixel, the smearing due to the PSF is only
marginal. The upper cut-off of FWHM=10~pixel used in this work, 
corresponds to a PSF-corrected physical size of 8.6~pc (core radius 4.3 pc).

%%%%%%%%%%%%%%%%%%%%%%%%%%%%% On Saturation %%%%%%%%%%%%%%%%%%%%%%%%%%%%%%%%%%%
\begin{figure}
    \centering
	\hspace*{-1.0cm}
    \includegraphics[width=95mm]{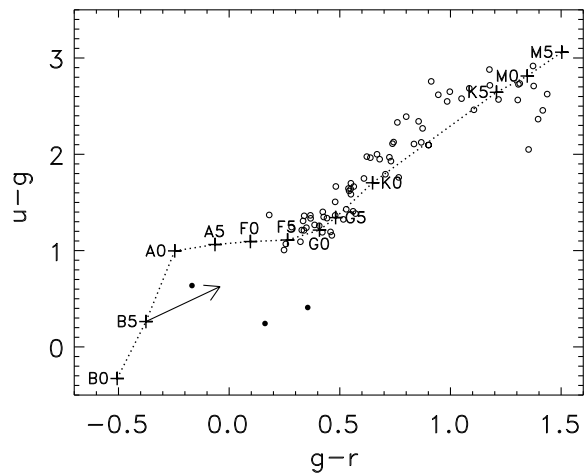}%{Fig2}
    \caption{The SDSS $u-g$ vs. $g-r$ diagram for all the objects that are
saturated in at least one of the HST/ACS bands. The locus of the main-sequence stars,
as well the direction of the reddening vector for $A_v=1$~mag \citep{Cardelli1989} are shown.
Only 3 objects could be interpreted as reddened young clusters (colours
bluer than $B5$ stars; shown by filled circles), with the rest following the track defined by the 
main-sequence stars of spectral types later than $F$. 
}
 \label{fig:sdss}
\end{figure}

We carried out a visual inspection of the images to make a list of objects
that have  diffraction spikes or are saturated in any one of the $B$, $V$ or $I$-band
images, with majority of them saturated only in the $I$-band.
A total of 83 such objects are found and there is no published information
about the nature of these objects from spectroscopic surveys such as the
one carried out by \citet{Sandage1984}, or several other follow-up studies.
We hence analyzed the colours of these   objects in order to investigate 
whether some of these could be compact young star clusters. 
To avoid the use of the {\it HST} colours that may be erroneous due to 
saturation, we carried out photometry of these objects using the Sloan 
Digital Sky Survey (SDSS) images and constructed a  $u-g$ vs. $g-r$ 
diagram for the selected objects,  which is shown in Figure~\ref{fig:sdss}.
The main-sequence colours are obtained using the \citet{Girardi2002} 
calculations for the grid of $T_{\rm eff}$ and $\log g$, that define each 
spectral type \citep{Mas1991}.
Notice that in this diagram the reddening vector is almost orthogonal to the
track defined by the spectral types for stars earlier than $A0$.
The colours of all except 3 objects are consistent with 
them being stars of spectral types later than $F$. These saturated objects
are extemely bright to be stars of M81, and hence, are likely to be foreground
Galactic stars. However some of these objects could be GCs.  
Young clusters are expected to have colours of stars earlier than 
spectral type $B5$, and there are only 3 candidates (shown by filled circles) 
that can be interpreted as dusty young clusters. None of these are brighter
than the brightest selected cluster. Thus, our selection criteria have
not eliminated possible bright clusters from our cluster sample.

%%%%%%%%%%%%%%%%%%%%%%%%%%%%%%%%%%%%%%%%%%%%%%%%%%%%%%%%%%%%%%%%%%%%%%%%%%%%%%%
%%%%%%%%%%%%%%%%%%%%%%%%%%%%%%%%%%%%%%%%%%%%%%%%%%%%%%%%%%%%%%%%%%%%%%%%%%%%%%%

The selection criteria described above resulted in a catalogue of 1123 
compact stellar cluster candidates. Artificial sources due to stellar 
blending are still present in our catalogue, with their fraction increasing 
systematically at fainter magnitudes. These contaminating sources are
relatively bluer in colour, limiting principally our capability of
detection of blue clusters. Hence, we restrict most of our analysis to 
$B=22$~mag for the blue clusters. For relatively redder clusters, 
contamination by stellar blending is not a serious limitation allowing us 
to retain them all the way to $B=24$~mag. After applying this colour-based 
selection criterion, which will be discussed again in \S4.2, we ended up 
with a list of 435 clusters. 
The catalogue of \citet{Chandar2001} contains 114 clusters up to a limiting
magnitude of $V=22$. Thus, our wide-field search has more than
tripled the number of clusters in M81.

The $B$ magnitude used in this work is $ISOMAG$ parameter calculated in 
SExtractor.  This parameter measures the magnitude by integrating the
background-subtracted counts in all the pixels that define the source.
Colours were obtained by  subtracting the magnitudes calculated within a fixed
aperture of diameter=20~pixels. Aperture corrections as suggested by
\citet{Sirianni2005} were applied to the magnitudes in each filter.
The method adopted by us to calculate colours ensures that the internal errors on
the colours are minimum. We estimated the errors on the magnitudes and
colours using the multiple observations of the same star as described below.

The 29 ACS fields used in this study offered around 10\% area overlap between 
the adjacent fields (see Fig.~1). The overlap region contained around 30,000 
stars. We used the two independent photometric measurements for the common
stars in each of the $B,V$ and $I$ bands to estimate
typical photometric errors on the magnitudes. As expected, errors
are found to be the least for bright stars (0.10~mag for $B<20$~mag), 
increasing systematically for fainter stars (0.20~mag for $B=24$~mag). Similar
errors were estimated in all the three filters. Errors on any two bands are
found to be uncorrelated, and hence errors on the colours were 
calculated by quadratically adding the errors on the magnitudes.
All the magnitudes and colours quoted in this work are on the standard
Vega system of magnitudes.

Our list contains 20 of the 41 spectroscopically confirmed GCs 
\citep{Schroder2002,PerelmuterBH1995}.
Among the missing objects, 8 are outside our field of view, and another 8
have stellar appearance (most are saturated) on the ACS images. These
objects could be very compact GCs. However, given that the galactic halo 
stars share the metallicities and radial velocities of the M81
GCs, these 8 objects are most likely to be galactic stars, rather than very 
compact GCs. These objects are listed in Table~\ref{tab:noGCs} along with
their observational properties. The remaining eight objects classified as
GCs do not satisfy one or the other of our selection criteria.
We also recover 53 of the 114 objects reported by \citet{Chandar2001b}. 
The principal reasons for the absence of the rest of the \citet{Chandar2001b} 
clusters are either they are blended stars (ellipticity$>0.1$) or that they
are foreground or M81 field stars wrongly classified as clusters due to the relatively 
poorer spatial resolution of WFPC2 images as compared to our ACS images. 
Thus, our catalogue of CSCs supersedes \citet{Chandar2001} catalogue,
both in its robustness of selection and in the spatial coverage.

\section{ANALYSIS}
\subsection{Colour Histogram}

\begin{figure}
    \centering
    \includegraphics[width=\columnwidth]{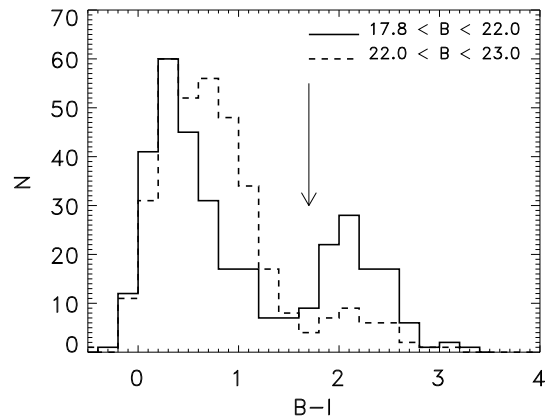}%{Fig2}
    %\vskip -3mm
    %\caption{The $B-I$ colour histogram for the CSC population is plotted 
    \caption{The  F435 $-$ F814 ($B-I$ for brevity) colour histogram for
the CSC population is plotted
      separately for the bright ($B<22$~mag) and relatively faint 
      ($B=$22--23~mag) clusters. Our CSC sample clearly divides into blue and
      red groups, with the dividing colour being $B-I=1.7$, which is shown 
       by the downward pointing arrow.}
 \label{fig:ColorHist}
\end{figure}

In Figure~\ref{fig:ColorHist}, we present the $B-I$ colour histograms for the 
cluster candidates separately for bright ($B<22$~mag; solid line) and faint 
($22.0<B<23.0$~mag; dotted line) members of our catalogue. It can be easily 
noticed that the distribution is bimodal in nature, especially for the 
brighter sample. Bimodality is also seen in the distribution of the $B-V$
colours for our sample objects. This bimodality has been noticed previously 
by~\citet{Chandar2001b}, who used this property to separate GC candidates
from the relatively younger clusters. In the next section, we use 
colour-magnitude diagrams to firmly establish this interpretation.
Based on this bimodality, we separated the cluster sample into two groups: a blue
group with $B-I<1.7$  and a red group with $B-I>1.7$. 
The colours of the red group members compare well with the colours of GCs 
in the Milky Way and M31 \citep{Harris1996}, and the colours of the blue 
clusters are similar to those of young and intermediate age clusters found in 
the Magellanic Clouds and M33 \citep{1999ApJ...517..668C}.
In Table~\ref{tab:catalog} we include the physical properties of the brightest 
compact clusters. 

\subsection{Colour-Magnitude Diagram}

\begin{figure*}%[t]
    \centering
    \includegraphics[width=1.40\columnwidth]{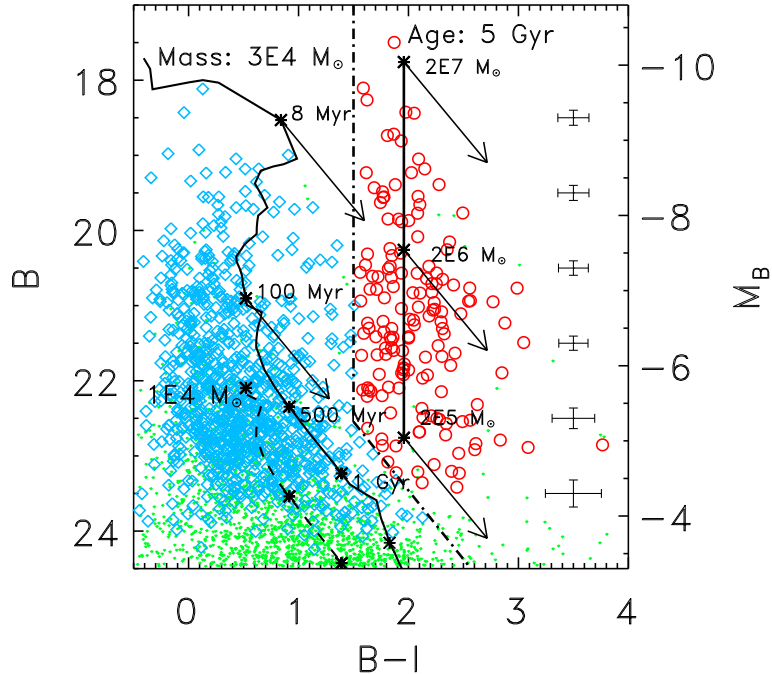}%{Fig3} %3
    %\includegraphics[width=1.40\columnwidth]{Fig3_Grafica_BI_B_color.ps}%{Fig3} %3
    %\vskip -4mm
    \caption{ $F435W-F814$ vs $F435W$  ($B-I$ vs $B$ for brevity) 
colour-magnitude diagram for the M81 CSC population: 
rhombus for the blue group clusters and circles for red group clusters.
The point-dashed line is the adopted line that separates the two groups.
Typical photometric errors on magnitudes and colours are shown at 
selected magnitudes on the right hand side. The small points in the bottom 
part of the figure represent point sources which are mainly field stars in M81. 
Only one out of 25 stars are plotted, for the sake of clarity.
Blue clusters fainter than $B\sim22$~mag are severely contaminated by sources 
formed by blending of these stars. 
An evolutionary track for a solar metallicity ($Z=0.019$) SSP \citep{Girardi2002} of 
mass$= 3\times 10^{4}\,M_{\odot}$ and age ranging from 3 Myr to 2 Gyr is shown by the
thick line. Another SSP for a mass of $10^{4}\,M_{\odot}$ is shown
between ages 0.1--1~Gyr by the dashed line. Note that majority of blue clusters
lie between these two SSPs.
A $Z=0.008$ SSP of constant age of 5 Gyr, but with masses in the range of 
$2\times~10^{5}$--$2\times 10^{7}\,M_{\odot}$ is also shown. It fits the colours and 
magnitudes for the red group which are globular cluster candidates. Reddening 
vectors corresponding to a robust upper limit of $Av=1$ mag are shown at 
selected positions on the tracks.
All the observational points are corrected for the foreground Galactic 
reddening using $E(B-V)=0.08$~mag.
}
\label{fig:CDM}
\end{figure*}

Compact stellar clusters are the closest observational analogs of the 
theoretically defined simple stellar population (SSP), where all the
stars are assumed to form in a single burst. The age and mass of the CSCs
can be obtained by comparing their observed colours and magnitudes with the 
corresponding quantities from a model SSP, in a Colour-Magnitude Diagram (CMD).
In Figure~\ref{fig:CDM}, we present the CMD, where all the clusters 
in our sample are plotted, with the blue clusters denoted by rhombus symbols, 
and the red clusters by circles.
Typical photometric errors on colours and magnitudes for point sources are shown
on the right-hand side of the plot. 
The errors on both the magnitudes and colours are expected to be
slightly larger for extended objects, such as CSCs. 
%The observed points are shown without correction for any foreground extinction. 
Two SSPs of cluster masses  3$\times10^4$~\msol\ and $10^4$~\msol\ 
are shown superposed on the blue group, between the age ranges of 3~Myr to
2~Gyr, and 0.1--1~Gyr, respectively. For the red
group, we show the locus of points for various cluster masses, all of age 
of 5~Gyr. The reddening vectors corresponding to $Av=1$~mag are shown at 
selected points on the SSP.
Location of point sources in the CMD is shown by the dots, which occupy 
mainly the bottom-left part of the diagram. In order to avoid agglomeration
of points in the plot, we show only 1 out of 25 stars. Extended sources formed 
by blending of these point sources are the principal contaminants of our 
cluster sample at magnitudes fainter than $B=22$. 

The plotted SSPs correspond to \citet{Girardi2002} solar metalicity (z=0.019) 
models for the blue clusters and $Z=0.008$ models for the red clusters. 
In their models, the colours and magnitudes were calculated directly 
for the HST/ACS filters and hence there was no necessity of converting
the magnitudes into standard Cousins-Johnson filter system.
The models assume a Kroupa initial mass function [IMF, \citet{Kroupa2001}] 
\footnote{The public distribution of SSPs uses a slightly modified IMF. We
multiplied the masses by a factor of 2.5, as suggested by the authors, to
bring it to the standard Kroupa IMF
}.

The brightest blue and red clusters of our sample have $B=18.45$~mag and
$B=17.83$, respectively. 
In the field of view of M81, there are 83 compact objects brighter than 
these limits. But, from the analysis of their SDSS colours (see \S3), we found 
that only 3 of the 83 objects have $u-g$ colours consistent with those of blue 
clusters, with none of these 3 brighter than the brightest blue cluster, even 
after taking into account possible reddening.
However the possibility exists, that some of the 
stellar-like objects are compact GCs in M81.

The difference in colour between the blue and red group clusters is more 
than a magnitude for the brightest clusters, with the separation gradually 
decreasing at fainter magnitudes. 
Before discussing the SSP ages of these populations, we first discuss the
mean amount of reddening expected towards the clusters.
The foreground Galactic reddening towards M81 is $E(B-V)=0.08$ mag
\citep{Schlegel1998}.
\citet{Kong2000} used photometry in 13 bands to map the reddening in 
spatial scales of 1.7$^{\prime\prime}$. 
The reddening values they measured in the bulge region was comparable to 
that expected for the foreground Galactic reddening,
indicating the absence of dust there. They
estimated a mean reddening of $E(B-V)=0.2$~mag ($Av=0.6$~mag 
using $Av/E(B-V)=3.1$) for the disk, including in the spiral arms. 
However, at the spatial scales of CSCs (few parsecs), reddening could be
different, and there are no such measurements for M81. Moreover, one may expect
higher reddening values when the clusters are young,
with the reddening decreasing once the clusters move out of, or destroy, their
parental clouds \citep{Bastian2005,Mengel2005}. Clusters in the blue group are associated with the spiral
arms (see section 4.3), and hence we cannot rule out them being very young
clusters. For example, most of the clusters with $B=$19--20~mag, could
be reddened young (age$<3$~Myr) clusters of mass $3\times10^4$~\msol\ for
a reddening value of $E(B-V)=0.5$~mag. A visual examination of these clusters
indicates that there are dusty features around some of them.

Taking into account the small expected reddening, 
the plotted SSPs for the masses  3$\times10^4$~\msol\ and
$10^4$~\msol\ represent well the observed
distribution of blue group clusters with $B<22$ mag in the CMD. 
The brightest clusters are expected to be the youngest in a normal star-forming
galaxy such as M81 \citep{Bastian2008}. 
In such a scenario, the masses for
the brightest two clusters would be between 1--2$\times10^4$~\msol.
The remaining clusters that lie within the 1$\sigma$ observational error from
the SSP could be objects with masses marginally above $10^4$~\msol,
with their ages ranging between 3~Myr and $\sim100$~Myr.
The ages could reach up to 300 Myr if their masses are less than 10$^4$\msol. 
At fainter magnitudes ($B>22$~mag) there are several clusters bluer than 
the SSP which share the same colour range as that for individual stars. 
Many of these sources may not be clusters and instead blended stars and hence
we exclude them from detailed analysis. 

While the observed distribution of blue group clusters closely follows the
plotted evolutionary track for mass$=3\times10^4$~\msol, the spread in
$B-I$ colour is clearly more than that can be accounted by the observational errors. After around
10~Myr, model colour lies between 0.6--0.8 up to around 300~Myr, whereas
most of the clusters brighter than $B=22$~mag have $B-I$ colour between 
0--1~mag, with the distribution peaking at $B-I=0.25$~mag (see
Figure~\ref{fig:ColorHist}). The scatter
on the redder side can be understood in terms of interstellar reddening.
On the other hand, the blueward shift of the distribution with respect to the
colour expected for the 10--300~Myr SSP could be due to the 
stochastic sampling of the stellar IMF that affects the colours
of low-mass clusters \citep{Cer2004, Maiz2009}.  
%combination of the following
%two cases: (1) most of the clusters have masses $<3\times10^4$~\msol\ and 
%they are in their very young phase (age$<5$~Myr), and (2) stochastic 
%sampling of the stellar IMF affects the colours of low mass clusters as 
%shown by \citet{Cer2004}. 
Hence, the presence of many clusters bluer than $B-I<0.5$~mag suggests
that majority of the clusters has mass of $<10^4$~\msol.
%%%DONE

The clusters in the red group are consistent with an SSP of age between
2--12~Gyr, with the mean colour of the group corresponding to an age of 5~Gyr.
The distribution of the $B$ magnitudes is entirely caused by the distribution
of masses of these clusters. Observed range of $B$ magnitudes corresponds to
a mass range of $10^{5}- 2\times10^{7}\,M_{\odot}$. Notice that a change in 
the age by a factor of 2 implies a change in the mass by also a factor of 2 
in this range of ages. The derived range of masses are similar to the values 
expected for globular clusters~\citep{Harris1996}. It may be noted that 20 
of our objects have been spectroscopically confirmed as globular clusters. 
Thus it is very likely that most of the red group objects are globular clusters.

The above analysis has established the blue and red groups as two distinct
families, the former belonging to the class of Super Stellar Clusters 
seen in starburst environments, and the latter being globular clusters.
There is a possibility, though very unlikely, that both belong to a single
family, and extinction separates them artificially into blue and red groups. 
For this to happen, all red group clusters should be experiencing an extinction
$Av\sim1.5$~mag. This would imply that the red clusters are preferentially 
associated to high extinction zones such as the spiral arms.
The discussion in the next section, where we analyze the spatial distribution
of the clusters in these two groups, clearly discards such a possibility.

\subsection{Spatial Distribution}\label{sec:spatdist}

\begin{figure}
 \centering
  \includegraphics[width=\columnwidth]{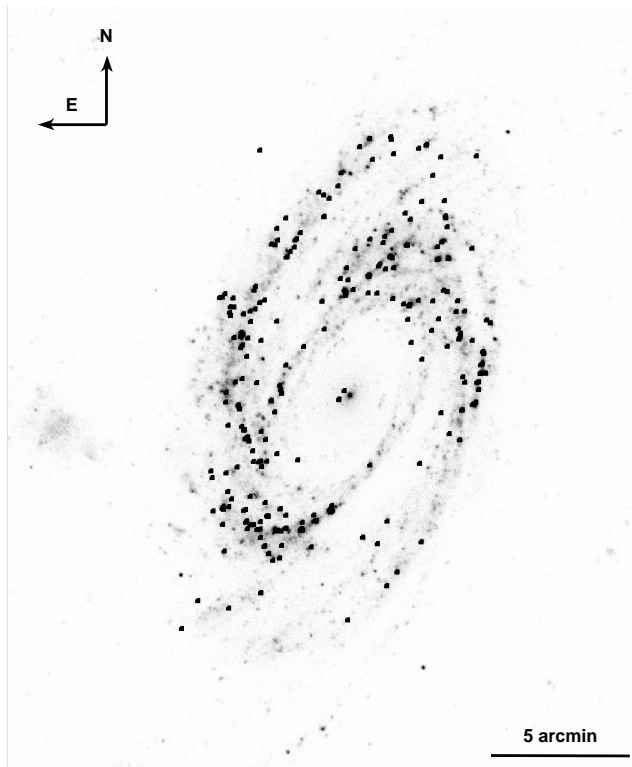}%{Fig4}
  %\vskip -3mm
  \caption{Spatial distribution of clusters in the blue group, superposed
on the \textit{GALEX} near UV image. It is clear that the blue clusters 
trace the spiral arms of M81.
    }
  \label{fig:blue}
\end{figure}

The spatial distribution of the blue clusters brighter than $B=22$~mag 
is presented in Figure~\ref{fig:blue}, where the gray scale image is 
the near UV image obtained from the GALEX archive
\footnote{http://galex.stsci.edu/GR4/}. 
The UV image traces star forming regions younger than 1 Gyr and delineate 
the position of the spiral arms in M81 \citep{2005ApJ...619L...1M}. 
The young clusters are located mainly on top of the spiral arms, suggesting
that they are  spatially and kinematically related to the population producing the UV emission.

\begin{figure}
 \centering
  \includegraphics[width=\columnwidth]{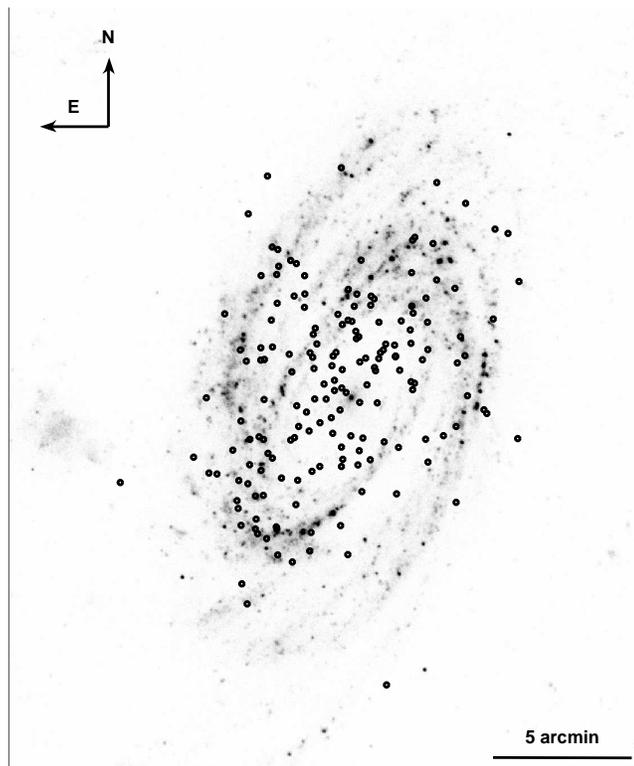}%{Fig5}
  %\vskip -3mm
  \caption{ Spatial distribution of clusters in the red group, superposed
on the \textit{GALEX} near UV image. The red clusters are uniformally 
distributed over the face of M81.  }
  \label{fig:red}
\end{figure}

On the other hand, the red clusters are homogeneously distributed 
over the face of M81 (see Figure~\ref{fig:red}). Most of the clusters are
seen at small radii, superposed on the bulge, with their number decreasing 
rapidly in the outer parts. It may be recalled that the bulge region hardly
suffers any interstellar extinction \citep{Kong2000}, and hence
the association of the red clusters with low extinction zones 
again re-iterates the idea that reddening is not the reason 
for their red colours, instead they are globular clusters.


\subsection{Luminosity Functions}

\subsubsection{Blue group clusters}

\begin{figure}%[!h]
    \centering
    \includegraphics[width=1.\columnwidth]{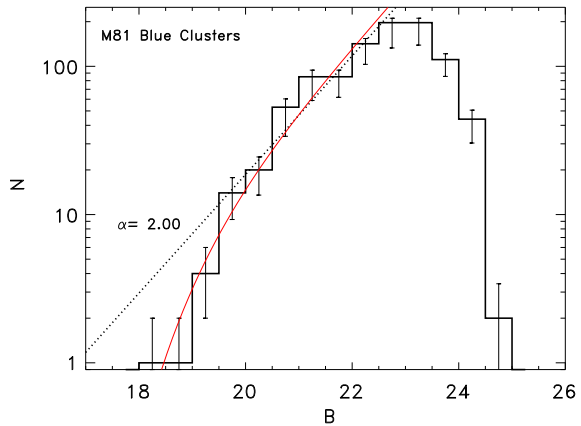}%{Fig6}
    %\vskip -3mm
    \caption{The LF of the blue group  clusters (histogram). A power-law 
function of $\alpha=2.0$ (dotted line) and the best-fit Schechter
function (solid line) are also plotted. 
Statistical errors calculated as $\sqrt{N}$ are shown.}
 \label{fig:YoungLF}
\end{figure}

A histogram of the  $B$-band luminosity function (LF) of the clusters of the blue group 
is shown in Figure~\ref{fig:YoungLF}. In the magnitude
range B=19--22 mag, the histogram follows a power
law of index $\alpha=2.0$ (dotted line). 
An un-biased fitting method which makes use of variable bin 
sizes~\citep{2005Maiz} gives a value for the power law index $\alpha=1.952\pm0.104$.
The histogram shows a peak at around 23~mag, 
deviating systematically from the power-law at fainter magnitudes. 
This decrease could be due to an intrinsic drop of the number of low 
luminosity (and mass) clusters. However, incompleteness of the sample at 
$B>23$~mag contributes significantly to the turnover. 
The incompleteness originates due to one of the selection criteria we have
imposed which requires selected clusters to have a minimum AREA of 50~pixels. 
Compact clusters ($FWHM<5$ pixels) fainter than $B>23$~mag do not satisfy 
this criteria, and hence are missing from our cluster sample.

A power-law index of $\alpha=2.0$, obtained in this study is the canonical 
value found in young stellar clusters in many starburst galaxies
\citep{Elson1985, deGrijs2003}. 
On the brighter end of the  $B$-band LF, the observed number of clusters is systematically
smaller than that expected for a power-law of index 2 ---
the sample contains six blue clusters brighter
than B=19.5~mag, whereas 18 such clusters are expected for a power-law of
index $\alpha=2.0$. Significantly, there are no blue clusters brighter
than $B=18.45$~mag (i.e. mass$ ~3.0\times 10^{4}M_{\odot}$; see Figure~\ref{fig:CDM}).
\citet{Gieles2009} has found that a Schechter function fits better the observational
data than a power-law function in the whole range of cluster luminosities.
We show in Figure~\ref{fig:YoungLF} that a Schechter function fits our data
also very well. The characteristic luminosity of the best-fitting function is
$B=19$~mag, which is more than 2 mag fainter than that expected for the
characteristic mass obtained by \citet{Gieles2009}. This issue is discussed
more in detail later in this section.

\begin{figure}%[!h]
    \centering
    \includegraphics[width=1.\columnwidth]{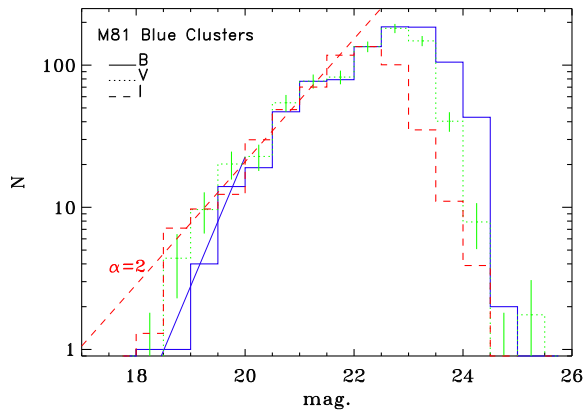}%{Fig7}
    %\vskip -3mm
    \caption{Comparison of the LFs of the blue group clusters (histogram) in $B$, $V$ and $I$ bands. 
A power-law function of $\alpha=2.0$ (dotted line) is shown, which fits very well the $I$-band 
data over the range 18--22~mag. On the other hand, $B$-band data shows steepening at magnitudes
brighter than $B=20$~mag.
}
 \label{fig:YoungHist3colors}
\end{figure}

\citet{Gieles2009} studied the steepening of the power-law as a function
of wavelength, and found that the function steepens more at longer
wavelengths as expected for Schechter functions.
This tendency for steepening of the luminosity function at high luminosities (or equivalently mass)
has been noted in several recent studies \citep{Haas2008, Larsen2009}.
Given the relatively small number of clusters in our M81 sample,
we could reliably obtain slopes in only two magnitude ranges ---
the dividing magnitude being 20. The resulting piece-wise linear fit
to the B-band data is shown in Figure~\ref{fig:YoungHist3colors}, 
where we can clearly see that the brighter-end slope is steeper.  
In the other two bands, a power law with $\alpha=2$ fits well the entire
range. Thus, we don't see the trend reported by~\citet{Gieles2009} for M51
in our dataset for M81. The extinction seems to be the reason for the
steepening at the bright end. We examined the HST image around the 10
brightest $I$-band clusters, and found traces of dust lanes running close
to these clusters, which  agrees with the above interpretation.
Hence, the conclusions derived from only the $B$-band LF could be misleading 
unless it is confronted with the $I$-band LF.

\begin{figure}%[!h]
    \centering
    \includegraphics[width=1.\columnwidth]{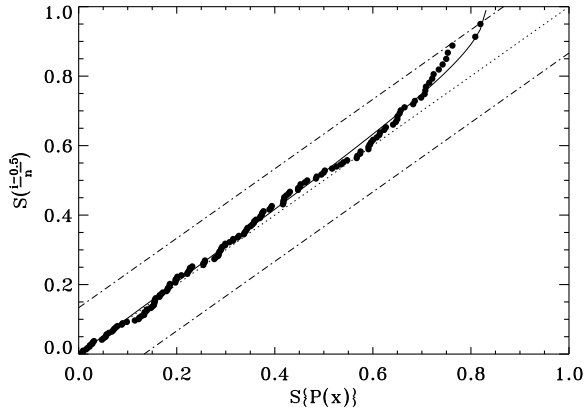}%{Fig7}
    %\vskip -3mm
    \caption{
The resulting stabilized percentile-percentile plot for
the I-band luminosity function. Two different null hypothesis have
been plotted, an infinite power law (dotted line) and a truncated
power law with index of $\alpha$ = 2.2 and an upper limit of 18.0 mag.
(solid line). Clearly the observed data (solid symbols) supports
the truncated power law hypothesis. The acceptance region of
the Kolmogorov-Smirnov statistic (significance level 5 per cent)
is given by the two parallels to the diagonal.
}
 \label{fig:truncation}
\end{figure}
Whereas the B-band LF can be fitted with a
Schechter function, the I-band LF shows signature
of a truncated power-law. We carried out the statistical
tests suggested by~\cite{2009Maschberger}
and firmly established the existence of a truncation
in the LF as can be seen in Figure~\ref{fig:truncation}. The modified
maximum likelihood method gives us an index of
$\alpha=2.219\pm0.130$ and an upper limit of 18.0~mag. The
graphical inspection of the data through the stabilized
percentile-percentile plot (Figure~\ref{fig:truncation}) and the different
goodness of fit test applied to the data confirm
the nature of the observed LF\footnote{We carried out the 10 different tests 
suggested by \cite{2009Maschberger}, and all of them are consistent
with the observed distribution being a truncated PL.}.
The observed truncation in LF corresponds to a mass of $2.4\times10^4 M_{\odot}$
for clusters of 7~Myr age, and $3.8\times10^4 M_{\odot}$ for 10~Myr age.
If the clusters are younger than 7~Myr, the truncation 
mass would be even less. This mass is an order of magnitude lower as
compared to the universal characteristic mass suggested by 
\citet{Larsen2009}.  
Only if all the brightest clusters are as old as 100~Myr, the 
truncation mass in M81 would correspond to the values 
obtained by \citet{Larsen2009}. In general, the brightest clusters in normal 
star-forming galaxies such as M81 are also the youngest 
(Bastian 2008).
Hence, it is very unlikely that bright clusters in M81 are as old as 100~Myr.
Thus, the inescapable conclusion from the above analysis is that the cluster
data of M81 are not consistent with the idea that the characteristic mass
of the LF is universal at $\sim2\times10^5 M_{\odot}$.

The lower truncation mass in M81 can also be infered by the relation between
the absolute magnitude of the brightest cluster, $M_V$(brightest) vs.
the SFR of the host galaxy, as is illustrated by \citet{Bastian2008}.
For the currently observed SFR of 0.75~M$\odot$\,yr$^{-1}$
\citep{Kara2007} in M81, the expected
$M_V$(brightest)$=-11$~mag, whereas the observed brightest cluster has $M_V=-9$~mag,
which lies about $2\sigma$ below the relation found in \citet{Bastian2008}.
Such a low truncation mass is also observed in M31 \citep{2009ApJ...703.1872V}, 
which is another normal star-forming galaxy like M81.

The interaction between M81 and its neighbour M82 
had triggered a disk-wide starburst in M82 around 300~Myr ago, that formed a rich 
population of compact clusters in the disk of M82 \citep{Mayya2006,Mayya2008}. 
It is very likely that the population of intermediate age clusters in M81 
is also formed following the same interaction event, and hence are coeval 
with the cluster population of M82. So, it is interesting to compare the 
luminosity functions of the cluster populations of these two galaxies. 
Though the two galaxies are at the same distance, 
the vast difference in extinction between the two galaxies ($<0.6$~mag in
M81 as compared to 1--6~mag in M82), inhibits a direct comparison of
the observed LFs. In M82, a power-law index of $\alpha=1.5$ was obtained
for the mass function, which is expected to be the index of the luminosity
function. Thus, observed indices of the  intermediate-age populations are 
distinctly different in M81 and M82. 
Another important difference in the cluster populations in these two galaxies
is the mass range of clusters --- there are no clusters massive than 
$\sim3\times10^{4}M_{\odot}$ in M81, whereas all the reported 393
intermediate-age clusters of M82 have masses $>2\times10^{4}M_{\odot}$.
The low value for the truncation mass observed in M81 is the reason for
the absence of massive clusters.

A possible reason for the absence of high mass clusters in M81 is that
it is a normal galaxy, whereas objects analyzed by \citet{Bastian2008}
are either starburst galaxies or small galaxies with
localized star forming sites. As discussed by \citet{Ashman2001}, the
formation of high-mass compact clusters
requires very high density gas, which is possible only
if the available gas mass is concentrated locally in regions of few
tens of parsecs. Such a condition can be easily satisfied in
starburst galaxies/regions such as studied by \citet{Bastian2008}, but not in
normal giant galaxies such as M81, where the star-forming sites (clusters)
are distributed throughout its large disk, as is illustrated in our Figure~\ref{fig:blue}.
The specific star formation rate -- defined as the SFR per unit of mass --
could play an important role in the formation of the most massive clusters observed in
a galaxy.
Therefore, it will be interesting to extend the study of the cluster population to more quiescent galaxies
in order to establish whether the $M_V$(brightest) depends on the  global SFR
or on the specific SFR. 

\subsubsection{Red group clusters}

\begin{figure}%[!h]
    \centering
    \small
    \includegraphics[width=1.\columnwidth]{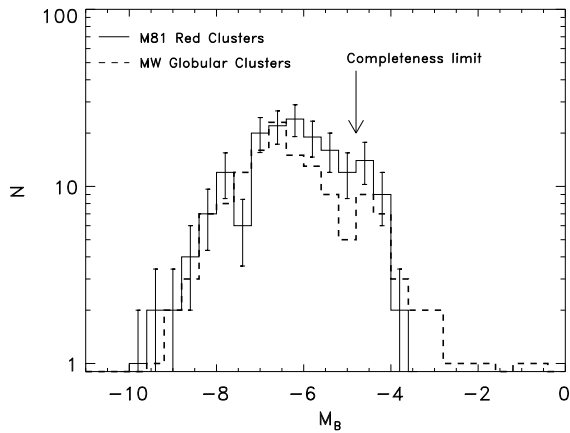}%{Fig8}
    %\vskip -3mm
    \caption{The LF of globular clusters detected in M81 (histogram) and the
      Milky Way (dashed line). 
Statistical errors calculated as 
$\sqrt{N}$ are shown and the completeness limit of our detection is also
marked with a downward pointing arrow.}
 \label{fig:OldLF}
\end{figure}

The LF for the red group clusters is plotted in Figure~\ref{fig:OldLF} along with that for
the Milky Way globular clusters from \citet{Harris1996}. It can be seen that the two 
distributions agree very well for $M_{B} < -4.5$ mag ($B\sim23$~mag), which is the completeness 
limit of our observations. The turnover in the distribution of Milky Way globular clusters 
at $M_{B}\sim-6.7$ [$M_{V}\sim-7.5$, \citet{Harris1996}] is also present in the distribution 
of the red group clusters of M81. Notice that, we have detected 172 globular clusters, 
which compares well with the 146 globular clusters in the Milky Way. It is heartening to find 
that the numbers and the LF between M81 and the Milky Way coincide, in spite of the complexity 
involved in the selection process. 

Our dataset allows us to compare the basic observational parameters of the 
brightest GC in M81 (R05R06584; Table~3) to the corresponding ones in other 
nearby galaxies, in particular those in the Milky Way ($\omega$~Cen) and M31 
(G1). Its colours are marginally bluer as
compared to the typical MW GCs, and suggest an age of $<5$~Gyr. 
Its absolute $F435W$-band magnitude corrected for Galactic extinction is 
$M_B^0=-10.3$~mag, which is $\sim0.9$~mag brighter than $\omega$~Cen, and
$\sim0.3$~mag brighter than G1 \citep{Meylan2001}. We derive a photometric mass 
of $2.5\times10^7$~\msol, using the \citet{Girardi2002}  Z=0.008 metallicity models. 
These derived properties indicate that the brightest GC in M81 is the most 
massive among the nearby galaxies. Follow-up spectroscopic studies of this
object will be invaluable in refining its mass and metallicity.

\section{DISCUSSION AND CONCLUSIONS}

Thanks to the superb spatial resolution of the \textit{ACS} camera on board
the \textit{HST}, 
we were able to obtain the largest sample of CSCs in M81 until now. We found a
total population of 435 CSCs brighter than $B=22$~mag, which increases by a 
factor of three the number of M81 CSCs catalogued previously 
by~\citet{Chandar2001}. The sample is divided into two well-defined 
populations: (a) a blue group with ages $<$300~Myr, 
masses $\sim~10^{4}\,M_{\odot}$ and distributed along the spiral arms of M81, 
and (b) a red group with ages 2--12~Gyr, masses between $10^{5}$ and
$2\times 10^{7}\,M_{\odot}$ that are distributed uniformly across the face of M81. 

Multi-band photometric and spectroscopic work on the M82 cluster population
seems to strongly favour an age of the M81/M82 interaction at 200-300 Myr.
These ages are in agreement with the results from numerical simulations
\citep{Yun1994, Yun1999}.
Hence, it is very likely that the same interaction also triggered the formation of
blue clusters in M81 as speculated by \citet{Chandar2001}.
However, it is important to remember that a firm conclusion on this
will have to wait for the spectroscopic confirmation of the ages.
Moreover, there is a large number of clusters below our confusion limit of $B=22$~mag,
which could have been formed before the interaction event (age$>300$~Myr; Figure~\ref{fig:CDM}) %Figure~\ref{fig:CMD}) 
if they are more massive than $\sim10^4$~\msol. If the age of this population
is confirmed to be older than $\sim300$~Myr, then these faint clusters could be part
of the normal star formation occurring in the disk of M81. In such a case,
the observed population of $B<22$~mag 
clusters discussed throughout this paper could as well be due to the normal star 
formation occurring in the disk of M81 independent of the interaction.

The $I$-band luminosity function of young clusters follows a power-law distribution
with an index $\alpha=2.21\pm0.13$.  
However, the commonly used $B$-band LF is better fitted with a double power
law or a Schechter function with a characteristic magnitude of $M_B=-9$. 
After careful examination of the $I$-band brightest objects,
which are probably the youngest, we conclude that the 
difference between the LFs in the I and B bands arises due to systematically 
higher extinction towards bright regions, thus affecting  the  high end of the 
LF of the B-band. 
The $I$-band LF truncates at the bright-end at $I=18.0$~mag ($M_I=-9.8$), 
which corresponds to a mass of $<4\times10^{4}\,M_{\odot}$, if the brightest
clusters are younger than 10~Myr. 
Thus, there is a clear absence of massive clusters in M81 as compared to those 
observed in M82 \citep{Mayya2008}  and other starburst galaxies \citep{Bastian2008}.
Models advocating universal characteristic masses of $2\times10^{5}\,M_{\odot}$
are inconsistent with the infered low truncation mass in M81. 

M81 is comparable to the Milky Way in its mass and morphology. GCs provide
a means to investigate the early formation history of galaxies. 
We find that the total number of GCs as well as their luminosity distribution
in M81 is very similar to that for the Milky Way. Thus, both these galaxies 
had very similar formation histories. 
The close encounter of M82 with M81 possibly created a new generation of 
compact clusters in the disk but did not affect the distribution of old
clusters that were in place at the time of the encounter.

\section{ACKNOWLEDGMENTS}
We would like to thank the Hubble Heritage Team at the Space Telescope Science
Institute for making the M81 images publicly available. 
We thank Thomas Maschberger and Pavel Kroupa for making the statistical
codes used in this work available through a dedicated web page.
We acknowledge the contribution to this work in the form of discussions 
with our colleagues Luis Carrasco, Lino Rodriguez, Abraham Luna and Olga 
Vega throughout the duration of the project.
This work was supported by CONACyT (M\'exico)
fellowship grants 4464 and project numbers 58956-F and 49942-F.
The authors are very grateful to an anonymous referee whose
comments and suggestions largely improved the clarity of this paper.

\label{lastpage}
\end{document}